\documentclass[reprint, amsmath,amssymb,aps, preprintnumbers, nofootinbib]{revtex4-1}
\usepackage{graphicx}				
\usepackage{amssymb}
\usepackage{amsmath, amsthm, amssymb}
\usepackage{dcolumn}
\usepackage{bm}
\usepackage{slashed}
\usepackage{tikz}
\usepackage{rotating}
\usepackage{hyperref}
\usepackage{array}
\usepackage{color}
\usepackage{multirow}
\usepackage{xcolor}
\usepackage[T1]{fontenc}
\usepackage{tocloft}
\usepackage[normalem]{ulem}
\usepackage{subcaption}
\usepackage{enumitem}
\usepackage{extarrows}
\usepackage{mathtools}
\usepackage{lipsum}
\usepackage{bbold}

\begin{document}

\preprint{DESY 19-147}

\title{FCNC-free multi-Higgs-doublet models from broken family symmetries}

\author{Ivo de Medeiros Varzielas}
\email{ivo.de@tecnico.ulisboa.pt}
\affiliation{CFTP, Departamento de F\'{i}sica, Instituto Superior T\'{e}cnico, Universidade de Lisboa, Avenida Rovisco Pais 1, 1049 Lisboa, Portugal}
\author{Jim Talbert} 
\email{james.talbert@desy.de}
\affiliation{Theory Group, Deutsches Elektronen-Synchrotron (DESY), Notkestra{\ss}e 85, 22607 Hamburg, Germany}

\begin{abstract}
We demonstrate how residual flavour symmetries, infrared signatures of symmetry breaking in complete models of flavour, can naturally forbid (or limit in a flavour specific way) flavour-changing neutral currents (FCNC) in multi-Higgs-doublet models (MHDM) without using mass hierarchies.  We first review how this model-independent mechanism can control the fermionic mixing patterns of the Standard Model, and then implement the symmetries in the Yukawa sector of MHDM, which allows us to intimately connect the predictivity of a given flavour model with its ability to sequester FCNC.  Finally, after discussing various subtleties of the approach, 
we sketch an $A_4$ toy model that realises an explicit example of these simplified constructions.
\end{abstract}
\maketitle

\section{Introduction}

Multi-Higgs-doublet models (MHDM) \cite{Lee:1973iz,Gunion:1989we,Branco:1999fs} are well motivated generalisations of the Standard Model (SM) with additional SU(2) doublet scalars.  For example, at least two Higgs doublets are required in supersymmetric extensions of the SM aiming to resolve the Hierarchy Problem, as well as in Peccei-Quinn solutions \cite{Peccei:1977hh} to the Strong CP Problem.  Additional scalars also introduce the possibility of new sources of CP violation beyond the SM (BSM), which will be necessary to explain the observed baryon asymmetry of the Universe \cite{Sakharov:1967dj}.

However, without additional assumptions, MHDM generically suffer from the fact that each Higgs doublet will couple to all families of fermions: up quarks, down quarks, charged leptons, and even neutrinos (if Dirac mass terms are added to the SM). The resulting proliferation of Yukawa couplings and neutral states associated with the extra scalars can potentially lead to large unobserved flavour-changing neutral currents (FCNC) at tree level. This is easiest to see in the Higgs basis, in which only one doublet has a non-vanishing vacuum expectation value (VEV) and thus all mass matrices are proportional to a single Yukawa. The mass matrix can be diagonalised for each family, but in general the Yukawa matrices associated with the other doublets remain arbitrary complex matrices.  Options to avoid this problem include
\begin{itemize}
\item {\bf{Natural Flavour Conservation}}, where fermions of a given electric charge can only couple to one type of Higgs doublet, thereby forbidding tree-level FCNC.  This can be enforced (and therefore remain radiatively stable) with Abelian discrete symmetries \cite{Glashow:1976nt,Paschos:1976ay,Weinberg:1976hu}.  For example, the famous Type-I, -II, -X, and -Y 2HDM fall within this classification,\footnote{For a thorough motivation and review of the 2HDM version of MHDM, see \cite{Branco:2011iw}.} as do the 3HDM and 4HDM extensions discussed in \cite{Cree:2011uy}.
\item {\bf{Mass Matrix Ans\"atze}}, where Yukawa matrices are assumed to have specific textures in flavour space, which when transforming to the basis of physical mass-eigenstates yields desirable SM mass and mixing phenomenology and sufficiently suppressed tree-level FCNC.  Attempts in this direction include \cite{Cheng:1987rs,DiazCruz:2004tr}. 
\item {\bf{Yukawa Alignment}}, where all Yukawa matrices $Y_{k}$ are perfectly aligned in flavour space (proportional to $Y_1$) and therefore simultaneously diagonalised, yielding no tree-level FCNC \cite{Pich:2009sp,Penuelas:2017ikk}.  Also see \cite{Rodejohann:2019izm}.
\end{itemize}
In this note we instead apply residual family symmetries (RFS), Abelian symmetries of the SM fermion mass sector, to the extended Yukawa sector of MHDM to show that they can easily forbid (or limit in a flavour-specific way) tree-level FCNC while simultaneously controlling the CKM and/or PMNS mixing matrices of the SM.  In the most predictive limit we recover a generalised form of Yukawa Alignment, which is well studied phenomenologically \cite{Penuelas:2017ikk}.  The RFS mechanism characterizes the symmetry breaking of a broad class of realistic flavour models, and is therefore a largely model-independent approach to controlling FCNC patterns in MHDM. Hence our study is distinct from previous attempts to texture MHDM Yukawas in symmetry-based effective models \cite{Serodio:2011hg, Varzielas:2011jr} or analyses implementing Abelian symmetries with unspecified dynamical origins \cite{Branco:1996bq,Ferreira:2010ir,Ivanov:2011ae,Serodio:2013gka,Ivanov:2013bka}.

The paper develops as follows:  in Section \ref{sec:REC} we review the RFS mechanism, highlighting the ability to connect observed patterns of fermionic mixing with UV flavour theories without specifying their explicit Lagrangians.  Then in Section \ref{sec:MHDM} we extend the RFS mechanism to the Yukawa sector of MHDM to achieve our core conclusions.  After discussing various subtleties we finally sketch an illustrative $A_4$ toy model in Section \ref{sec:MODEL} that realises a special case of the RFS formalism.  We conclude in Section \ref{sec:CONCLUDE}.

\section{Residual flavour symmetries}
\label{sec:REC}
The Yukawa sector of the SM exhibits accidental U(1)$^{3}$ symmetries corresponding to independent rephasings of each fermion generation. This can be seen directly from the Lagrangian in the broken phase, after electroweak symmetry breaking (EWSB), in the mass basis of the fermions:
\begin{equation}
\label{eq:SMyuk}
\mathcal{L}_{SM}^{Y} = -\left( \bar{l}^{i}_{L} \, m_{e}^{ij} \, e^{j}_{R} + \bar{u}^{i}_{L} \, m^{ij}_{u} \, u^{j}_{R} + \bar{d}^{i}_{L} \, m^{ij}_{d} \, d^{j}_{R} \right),
\end{equation}
with $\lbrace l, u, d \rbrace_L$ members of the former SU(2)$_L$ doublets, $\lbrace e, u, d \rbrace_R$ SU(2)$_L$ singlets, and where, due to our basis choice, $m^{ij}_{e,u,d}$ are diagonal matrices of mass eigenvalues for the charged leptons, up quarks, and down quarks, respectively.\footnote{For brevity we suppress the flavour indices $i,j$ from this point forward.}  Now let us apply transformations on the fermions of the form
\begin{align}
\nonumber
&A \rightarrow T_{A} A,\,\,\,\,\, \text{with} \,\,\,\,\, A \in \lbrace u_{L}, u_R, d_{L}, d_R, l_{L}, e_R\rbrace, \\
\label{eq:residualtransform}
 &T_{A} = \text{diag} \left(e^{i \alpha_{A}},e^{i \beta_{A}} , e^{i \gamma_{A}} \right).
\end{align} 
It is obvious that \eqref{eq:SMyuk} is invariant under the action of \eqref{eq:residualtransform} when the transformations on left- and right-handed fermions are equivalent, as they must be if fermion mass terms are generated.  While invariance under \eqref{eq:residualtransform} may seem trivial in the mass basis, when the same relation is rotated to the flavour basis, the transformations become functions of $U_{A}$, the (highly non-trivial) fermionic mixing matrices of the SM:
\begin{equation}
\label{eq:flavourbasis}
T_{A} \rightarrow T_{AU} = U_{A} \, T_{A} \, U_{A}^{\dagger}\,,
\end{equation}
which are appropriately interpreted as symmetries of the mass matrix (and not of the full $\mathcal{L}_{SM}$, as weak interactions do not respect them without additional restrictions):
\begin{equation}
\label{eq:flavourbasismass}
m_{AU} =  T^{\dagger}_{AU} \,m_{AU}\, T_{AU}.
\end{equation}
We consider scenarios where \eqref{eq:flavourbasismass} is not accidental, but is instead the remnant signature of an ultra-violet (UV) flavour theory controlled by a family symmetry $\mathcal{G}_{F}$ that commutes with the SM gauge group (or indeed any other BSM symmetry).  This can occur, for example, in models where the family-symmetry breaking occurs through the VEV of scalar flavons $\lbrace \phi \rbrace$. Regardless of the dynamics that yield it, the specific alignments of the VEV $\langle \phi \rangle$ in flavour space then generate SM Yukawa matrices that have the particular symmetries represented by $T_{A}$ embedded in them, such that
\begin{equation}
T_{A} \langle \phi \rangle_{A} = \langle \phi \rangle_{A}
\end{equation}
in a given fermion sector (and in the appropriate basis). The breaking of $\mathcal{G_F}$ can also arise e.g. from boundary conditions in models with extra dimensions. Hence the $T_{A}$ generate RFS if the mass matrices of the different fermion sectors are invariant under different subgroups $\mathcal{G}_{A}$ of $\mathcal{G}_{F}$. Schematically, an especially predictive symmetry breaking chain might go as
\begin{equation}
\label{eq:GF}
\mathcal{G_{F}}  \rightarrow \begin{cases}
				\mathcal{G_{L}}   \rightarrow \begin{cases} 
										\mathcal{G_{\nu}}
										\\
										\mathcal{G_{\text{l}}}
										\end{cases} \\
				\mathcal{G_{Q}} \rightarrow \begin{cases}
										\mathcal{G_{\text{u}}}
										\\
										\mathcal{G_{\text{d}}}\,,
										\end{cases}
				\end{cases}
\end{equation}
where $\mathcal{G}_{F}$ is of an unspecified mathematical nature --- it can be Abelian or non-Abelian, continuous or discrete.  However, if it is assumed to be discrete (as it would have to be if it is sourced from a larger discrete group, that may be non-Abelian), the RFS is instead taken to be a cyclic $\mathbb{Z}_{m}$ subgroup of U(1)$^3$, generated by a matrix representation $T_{A}$ whose phases are constrained such that $T_{A}^{m} = \mathbb{1}$.  


\begin{table}[t]
\renewcommand{\arraystretch}{1.5}
\noindent
\centering
\begin{tabular}{|c||c|c|c|c|c|c||c|c|c|c|}
\hline
\cite{Altarelli:2010gt} & $L_L$ & $\nu_R$ & $e_{R}$ & $\mu_{R}$ & $\tau_{R}$ & $H_{u,d}$ & $\phi_{l}$ & $\phi_{\nu}$ & $\xi$ & $\tilde{\xi}$\\
\hline 
SU(2)$_L$ & $\bold{2}$ & $\bold{1}$ &  $\bold{1}$ & $\bold{1}$ & $\bold{1}$ & $\bold{2}$ & $\bold{1}$ & $\bold{1}$ & $\bold{1}$ & $\bold{1}$  \\
\hline
$A_{4}$ & $\bold{3}$ & $\bold{3}$ & $\bold{1}$ & $\bold{1^{\prime \prime}}$ & $\bold{1^{\prime}}$ & $\bold{1}$ & $\bold{3}$ & $\bold{3}$ & $\bold{1}$ & $\bold{1}$ \\
\hline
\end{tabular}\, \\
\caption{The relevant field and symmetry content of the $A_4$ Altarelli-Feruglio model.  We have explicitly given the assignments in \cite{Altarelli:2010gt}.}
\label{tab:AF}
\end{table}


A well-known example where this occurs can be found in the supersymmetric Altarelli-Feruglio model of leptonic mass and mixing \cite{Altarelli:2005yx,Altarelli:2010gt}, whose relevant field and symmetry content is given in Table \ref{tab:AF}, where SM fields and new scalar flavons $\phi_{l,\nu}$, $\xi$, and $\tilde{\xi}$ are charged under a non-Abelian $A_4$ symmetry (to be identified with $\mathcal{G}_{\mathcal{L}}$).\footnote{Additional superfields and shaping symmetries are required for vacuum alignment, charged lepton mass hierarchies, and to prevent unwanted contact interactions in \cite{Altarelli:2005yx,Altarelli:2010gt}, but these are not required for understanding the appearance of RFS presently under discussion.  Also recall that in this model the charged leptons are already in a basis where their mass matrix is diagonal, so $U_{l} = \mathbb{1}$ and hence $U_{PMNS} = U_{\nu}$.}  These compose the leading-order effective super-potential,
\begin{align}
\nonumber
&\mathcal{L}_{AF} \supset y_e \left[\bar{L}_L \phi_l \right]e_R  + y_\mu \left[ \bar{L}_L \phi_l\right]^{\prime} \mu_R + y_\tau \left[\bar{L}_L \phi_l\right]^{\prime \prime}\tau_R \\
\label{eq:AFSP}
&+ y_\nu [\bar{L}_L \nu_R] + \left(x_A \xi + \tilde{x}_A \tilde{\xi}\right)\left[\nu_R \nu_R\right] + x_B \left[\phi_\nu \nu_R \nu_R \right],
\end{align}
where the bracket notation implies contractions to $A_4$ singlets, 
\begin{align}
\nonumber
&[ \bar{L}_L \phi_l ]\,\,\,= \bar{L}_{L1} \phi_{l 1} + \bar{L}_{L2} \phi_{l 3} + \bar{L}_{L3} \phi_{l 2},\\
\nonumber
&[ \bar{L}_L \phi_l ]' \,= \bar{L}_{L1} \phi_{l 2} + \bar{L}_{L2} \phi_{l 1} + \bar{L}_{L3} \phi_{l 3},\\
\nonumber
&[ \bar{L}_L \phi_l ]''= \bar{L}_{L1} \phi_{l 3} + \bar{L}_{L2} \phi_{l 2} + \bar{L}_{L3} \phi_{l 1}\,,
\end{align}
and where we have omitted factors of Higgs fields, irrelevant flavons, and mass suppressions (e.g. the term $y_e\left[\bar{L}_L\phi_l\right] e_R  \equiv y_e H^{\prime}_d \left[\bar{L}_L \phi_l \right] e_R \, \chi^4/\Lambda^5$ in \cite{Altarelli:2005yx,Altarelli:2010gt}, with $\chi$ a flavon associated to an additional Froggatt-Nielsen symmetry and $\Lambda$ the cutoff scale of the theory). Thanks to the model's scalar potential, the triplet flavons develop VEV aligned along
\begin{equation}
\langle \phi_l \rangle = \left(v_l, 0, 0 \right), \,\,\,\,\,\,\,\,\,
\langle \phi_\nu \rangle = \left(v_\nu, v_\nu, v_\nu \right),
\label{eq:flavonvev}
\end{equation} 
during flavour-symmetry breaking.  Then, after EWSB these preserve a $\mathbb{Z}_{3}$ and $\mathbb{Z}_{2}$ symmetry in the resulting charged lepton and Majorana neutrino mass terms respectively, with the latter obtained through a Type-I see-saw mechanism.  The $T$ matrices that generate these RFS are given by
\begin{align}
\label{eq:A4AF}
T_{l U} = 
\begin{pmatrix}
1 & 0 & 0 \\
0 & \omega & 0 \\
0 & 0 & \omega^2
\end{pmatrix},\,\,\, 
T_{\nu U} = \frac{1}{3}
\begin{pmatrix}
-1 & 2 & 2 \\
2 & -1 & 2 \\
2 & 2 & -1
\end{pmatrix},
\end{align}
where $\omega = \text{exp}(2 \pi i/3)$ and
\begin{equation}
\label{eq:A4AF2}
T_{\nu U} = U_{TBM} \cdot \,\underbrace{\text{diag}(-1,1,-1)}_{T_{\nu}} \cdot \,U_{TBM}^{\dagger},
\end{equation}
with $U_{TBM}$ the famous (albeit experimentally excluded) `tri-bimaximal' mixing matrix \cite{Harrison:2002er} predicting $\sin \theta_{12}^l =1/\sqrt{3}$, $\sin \theta_{23}^l = 1/\sqrt{2}$, and $\sin \theta_{13}^l = 0$.  Impressively, one recovers the parent symmetry of the theory by identifying the residual generators $T_{l,\nu}$ in the mass basis (as we did above) and rotating through its specific prediction for $U_{PMNS}$ --- it is easily checked that $A_{4}$ is the group closed by the matrices in \eqref{eq:A4AF}! 

From \eqref{eq:A4AF}-\eqref{eq:A4AF2} one observes that it is possible to use RFS to relate quark and/or lepton mixing 
to $\mathcal{G}_{F}$ in a model-independent way --- their information (and that of their analogues in other models, cf. \eqref{eq:residualtransform}-\eqref{eq:flavourbasis}) can be extracted from the low-energy Lagrangian without reference to the specific dynamics of any given BSM field and/or symmetry content.    Indeed, constructions embedding the symmetry breaking patterns in \eqref{eq:GF} (or relaxed versions of it) have been well studied in the context of fermionic mixing in the SM (see e.g. \cite{King:2013eh,Altarelli:2010gt,Grimus:2011fk} for reviews), with the literature including complete models realising the RFS as well as both analytic and computational approaches to cataloguing the types of $\mathcal{G}_{F}$ that can break to phenomenologically viable CKM/PMNS matrices \cite{Lam:2007qc, Ge:2011ih, Ge:2011qn, deAdelhartToorop:2011re, Hernandez:2012ra, Lam:2012ga, Holthausen:2012wt, Holthausen:2013vba, King:2013vna, Lavoura:2014kwa,Fonseca:2014koa,Hu:2014kca, Joshipura:2014pqa, Joshipura:2014qaa, Talbert:2014bda, Yao:2015dwa, King:2016pgv, Varzielas:2016zuo,Yao:2016zev,Lu:2016jit,Li:2017abz,Lu:2018oxc,Hagedorn:2018gpw,Lu:2019gqp}. 

\section{The Yukawa Sector of MHDM}
\label{sec:MHDM}
We now use the RFS mechanism as a tool to not only control fermionic mixing, but also FCNC in MHDM,\footnote{The concept can apply to other theories as well --- see the recent leptoquark application in \cite{deMedeirosVarzielas:2019lgb,Bernigaud:2019bfy} and note the correspondence between \eqref{eq:coreconstraint} and eq.(18) in \cite{deMedeirosVarzielas:2019lgb}.} where fermions generically couple to each of the $k$ Higgs doublets, yielding a Yukawa Lagrangian given by:\footnote{We will adopt the notation of \cite{Penuelas:2017ikk} in the following discussion.}
\begin{align}
\nonumber
\mathcal{L}^Y =  &- \sum^N_{k=1} \Big \lbrace \bar{Q}^{\prime}_L\,\left( Y^{d,\prime}_k \,H^{\prime}_k\, d^{\prime}_R + Y^{u,\prime}_k \,\tilde{H}^{\prime}_k\, u^{\prime}_R\right) \\
\label{eq:LYuk}
&+  \bar{L}^{\prime}_{L} \,Y^{e,\prime}_k \,H^{\prime}_k\, e^{\prime}_R + h.c. \Big \rbrace \,.
\end{align}
Here $Q_L$ and $L_L$ are SU(2)$_L$ doublets, $\tilde{H}_k = i \tau_2 H^\star_k$, and each Higgs is in the standard parameterization,
\begin{equation*}
H^{\prime}_k = \frac{e^{i\theta_k}}{\sqrt{2}}\left(
\begin{array}{c}
H^{+}_k \\
v_k + \rho_k + i \eta_k
\end{array}
\right)\,,
\end{equation*}
with $e^{i\theta_k} v_k /\sqrt{2}$ the neutral Higgs VEV whose global phase factor $\theta_k$ is defined such that $\theta_1 = 0$.  For now we have assumed a massless neutrino, and therefore there will be no physical mixing in the lepton sector, $U_{PMNS}=\mathbb{1}$.  We will allow for Majorana neutrinos and a non-trivial $U_{PMNS}$ in Section \ref{sec:MODEL}, but again these do not obtain their masses from the Higgs Mechanism.  Finally, the Yukawa matrices $Y^{\prime}_k$ in \eqref{eq:LYuk} are $3 \times 3$ matrices in flavour space, and after EWSB the associated mass matrix $M^{\prime}_k$ will receive contributions from the VEV of all $H^{\prime}_k$.

After EWSB we go, without loss of generality, to the Higgs basis where all the VEV is in $H_1$ (this is done by a Higgs-basis transformation, a unitary transformation on the $N$ doublets):
\begin{equation*}
H_1 = \frac{1}{\sqrt{2}}\left(
\begin{array}{c}
\sqrt{2}\, G^{+}\\
v + S^0_1 + i G^0
\end{array}
\right)\,, \,\,\,
H_{k>1} =
\frac{1}{\sqrt{2}}\left(
\begin{array}{c}
\sqrt{2}\,S^{+}_k \\
S^0_k + i P^0_k
\end{array}
\right),
\end{equation*}
where now the $G$ fields correspond to the electroweak Goldstone bosons.
We also go without loss of generality to the basis where the charged fermion mass matrices are diagonal (this is a unitary transformation on the flavour space of the fermions). In this mass basis, $Y^A_1$ is diagonal by definition  --- $H_1$ is the only doublet with a non-zero VEV and the mass matrix is necessarily proportional to $Y^A_1$.  On the other hand, $Y^A_{k>1}$ are still generically $3 \times 3$ matrices.  In this special basis (which we leave unprimed), \eqref{eq:LYuk} is then expanded as
\begin{align}
\nonumber
\mathcal{L}^Y = &- \left(1 + \frac{S^0_1}{v} \right) \left( \bar{d}_{L} m_d d_R + \bar{u}_L m_u u_R + \bar{l}_{L} m_e e_R \right) \\
\nonumber
&- \frac{1}{v} \sum_{k=2}^{N} \left(S^0_k + i P^0_k \right) \left( \bar{d}_{L} Y_k^d d_R + \bar{u}_L Y_k^u u_R + \bar{l}_{L} Y_k^e e_R \right)\\
\nonumber
&- \frac{\sqrt{2}}{v} \sum_{k=2}^{N} S^+_k \left( \bar{u}_L V Y^d_k d_R - \bar{u}_R Y_k^{u,\dagger} V d_{L} + \bar{\nu}_L Y^e_k e_R \right) \\
\label{eq:Lunprimed}
 &+ h.c.\,,
\end{align}
where $V \equiv U_{CKM}$, the second line is clearly capable of generating the dangerous FCNC we are concerned with, and the third line includes new charged-current (CC) interactions characteristic of the MHDM.  Note that the physical neutral (charged) scalar mass eigenstates $H^0_k$ ($H^+_k$) are linear combinations of the $\lbrace S^0_k, P^0_k \rbrace$ ($S^+_k$) field components of \eqref{eq:Lunprimed}, whose specific relations are controlled by the parameters of scalar potentials that our analysis does not require knowledge of. 
\subsection{RFS for MHDM}
In our formalism (assuming flavon-based spontaneous symmetry breaking) mass/Yukawa matrices originating from an effective operator inherit an invariance under the action of the RFS generators,
and hence each $m_A$/$Y^A_k$ in \eqref{eq:Lunprimed} is subject to constraints (for each $k$). 

Moving line-by-line we first observe that the mass term RFS analysis proceeds precisely as in the SM; as we are in the fermion mass-eigenstate basis we immediately conclude that non-degenerate fermion masses require $T_{A}$ to be in a diagonal representation, cf.\eqref{eq:residualtransform}, and that $T_{f_{L}} = T_{f_{R}}$.  

Armed with this insight one then concludes without loss of generality that, 
if the FCNC operators in the second line of \eqref{eq:Lunprimed} are to remain RFS invariant, the associated Yukawa couplings must respect the following RFS constraint:
\begin{equation}
\label{eq:RFSYukawa}
Y^A_k \overset{!}{=} T_A Y^A_k T_A^\dagger \,,
\end{equation}
where the $!$ notation implies $\emph{`must be equal to'}$.  For the $k \neq 1$ terms, expanding the implicit flavour indices in \eqref{eq:RFSYukawa} yields
\begin{widetext}
 \begin{equation}
 \label{eq:coreconstraint}
    \left(
\begin{array}{ccc}
 Y_{11} & e^{i(\alpha_{l}-\beta_{l})}\, Y_{12} &  e^{i(\alpha_{l}-\gamma_{l})}\,Y_{13}  \\
e^{i(\beta_{l}-\alpha_{l})}\,Y_{21}  & Y_{22}  & e^{i(\beta_{l}-\gamma_{l})}\, Y_{23}     \\
e^{i(\gamma_{l}-\alpha_{l})}\,Y_{31}  & e^{i(\gamma_{l}-\beta_{l})}\,Y_{32}   & Y_{33} 
\end{array}
\right)
  \overset{!}{=}
  \left(
\begin{array}{ccc}
Y_{11} & Y_{12} & Y_{13} \\
Y_{21} & Y_{22} & Y_{23} \\
Y_{31} & Y_{32} & Y_{33}  
\end{array}
\right)
\end{equation}
\end{widetext}
for each $k$.
Denoting a generic phase in $T_A$ by $\psi_A \in \lbrace \alpha_A, \beta_A, \gamma_A \rbrace$, we now make a few important observations:
\begin{enumerate}[label=\bf(A\arabic*)]
\item {\bf{FCNC:}}  Barring equalities amongst $\psi_A$, \eqref{eq:coreconstraint} forces each $Y_k$ to be diagonal in the same basis.  Tree-level FCNC are therefore trivially avoided.  We refer to this as the `strict' RFS limit, where we also observe that, in any family sector, the matrices $Y^{\prime}_k$ defined in the weak-eigenstate basis are simultaneously diagonalisable, since they undergo the same chiral rotations $U_A$ in transforming to $Y_k$. 

In the `weak' RFS limit where $\psi_A^i = \psi_A^j$, off-diagonal elements in the $(i,j)$ sectors of \emph{each} $Y_k$ are permitted and can induce FCNC.  However, the RFS cannot control the magnitude of the couplings, only their shape; experimental bounds on $\Delta F =1,2$ processes can be found in \cite{Tanabashi:2018oca}.
\item {\bf{Fermionic Mixing:}} If the RFS is to also control complete three-generation $U_{CKM,PMNS}$ mixing, there \emph{cannot} be any phase equalities in \eqref{eq:coreconstraint} unless multiple RFS are active (cf. \eqref{eq:productgroup}), due to the following (flavour-basis) equality:
\begin{equation}
T_{AU} = U_{A}\,T^{ii=jj}_{A} \, U_{A}^{\dagger} = U_{A}\,R_{A}^{ij} \,T^{ii=jj}_{A} \,R_{A}^{ji\star}\, U_{A}^{\dagger}.
\\
\label{eq:degenEV}
\end{equation}
Here $R$ is a $2\times2$ unitary rotation matrix in the $(i,j)$ plane given by
\begin{equation}
\label{eq:mixdegen}
R^{ij} \equiv 
\left(
\begin{array}{cc}
\cos \theta_{ij} & \sin \theta_{ij} \,e^{-i\delta_{ij}} \\
-\sin \theta_{ij} \,e^{i\delta_{ij}} & \cos \theta_{ij}
\end{array}
\right). 
\end{equation}
That is, the parent flavour symmetry $\mathcal{G_F}$ cannot distinguish the physical mixing matrix $U_A$ from $U_A\,R_{A}^{ij}$ if $\psi^i_{A} = \psi^j_{A}$ in the RFS generator $T_A$ --- the low energy model either permits free parameters or they are controlled by alternative mechanisms (e.g. accidental symmetries). 

All $Y_k$ (in all family sectors) are therefore simultaneously diagonal when a model realises RFS to completely control CKM and PMNS mixing.  Concurrently, when $Y_k$ have FCNC-inducing off-diagonal elements, they are in precisely the same matrix sector that the RFS cannot control in the CKM and/or PMNS.  

For example, the RFS could control the dominant Cabibbo mixing in the (1,2) sector of the CKM matrix while simultaneously protecting against the most dangerous FCNC from the light quark generations. 
\item{\bf{Charged-Current Interactions:}} As with the weak interactions of the SM, the novel CC in the third line of \eqref{eq:Lunprimed} are not RFS invariant despite the fact that, unlike in the SM, these interactions are sourced from the same flavon-enhanced operators as the mass terms, and so one might naively expect the RFS to hold.  However, since we take $\Lambda_F > \Lambda_{\text{EW}}$, the imprint of $\mathcal{G_F}$ through $\mathcal{G}_{A}$-symmetric $Y_k^A$  occurs without reference to the isospin decomposition that yields the different terms in \eqref{eq:Lunprimed} after EWSB.\footnote{This is easiest to see in models where Higgs fields transform as trivial singlets under $\mathcal{G_F}$, as we will sketch in Section \ref{sec:MODEL}.}  Hence the RFS is again properly understood as a symmetry of mass/Yukawa matrices (cf. \eqref{eq:flavourbasismass}) which controls $Y_k^A$ in the CC, but not the overall term.

Regardless, using that $Y_k^d$ is forced into a specific shape through \eqref{eq:coreconstraint}, we observe that one can artificially force the CC to be RFS invariant (for all $k$) via
\begin{equation}
\label{eq:CCrestrict}
T_{u}^{\dagger}\,V\,T_{d} \overset{!}{=} V \,\, \Longrightarrow V\,T_{d} = T_{u}\,V\,.
\end{equation}  
Intriguingly, \eqref{eq:CCrestrict} also leaves the SM CC mediated through $W^{\pm}$ RFS invariant, and so it can be imposed as an additional condition. However, we observe that \eqref{eq:CCrestrict} implies 
\begin{equation}
\label{eq:commute}
T_{uU} = T_{dU}\
\end{equation}
in the flavour basis.  Hence $\mathcal{G_{F,Q}}$ must be Abelian if it is closed via this generator alone, which is less phenomenologically interesting to engineer in flavon models.  If $T_{eU}$ is also part of the overall generating set then $\mathcal{G_F}$ can again be non-Abelian, and an equality with $T_{\nu U}$ analogous to \eqref{eq:commute} could appear if neutrino masses and non-trivial $U_{PMNS}$ are considered. 
\item{\bf{Product Group RFS:}}
We have implicitly assumed that $\mathcal{G}_A$ are generated by one matrix $T_A$.  If the RFS are instead described by product groups of the form
\begin{equation}
\label{eq:productgroup}
\mathcal{G}_A \cong \mathcal{G}_{A}^1 \times \mathcal{G}_{A}^2 \times ...
\end{equation}
then each RFS generator $T_A^i$ enforces equalities analogous to \eqref{eq:coreconstraint}, and so FCNC-inducing $(Y_k)_{ij}$ may be permitted by $\mathcal{G}_{A}^1$ but not by $\mathcal{G}_{A}^2$. While \eqref{eq:productgroup} can control any of the fermion families, we recall that the situation occurs naturally in the Majorana neutrino mass matrix, whose (maximal) RFS is well-known to be a Klein four-group, $\mathcal{G}_\nu \cong \mathbb{Z}_2 \times \mathbb{Z}_2$.  Note that \eqref{eq:productgroup} can also alleviate ambiguities in mixing predictions entering through \eqref{eq:degenEV}.
\item {\bf{Generator Representation:}} Applying the same $T_A$ representation in each Yukawa operator for all $k$ embeds an inherent (albeit natural) assumption about the structure of the family symmetry breaking in these terms.  In models where each operator is enhanced by flavon field insertion(s), it implies that the VEV $\langle \phi \rangle$ are aligned in flavour space such that the same RFS is preserved.\footnote{This does not mean that the VEV must be equivalent, however.}  In the event $\langle \phi _k \rangle$ does not preserve the same RFS as $\langle \phi_{k^{\prime}} \rangle $, e.g., a different $T_A$ should be applied in each term (which is of course possible).  In this more baroque scenario FCNC-inducing elements can then be in different matrix sectors. 
\item {\bf{Generalised Yukawa Alignment:}} In the strict RFS limit our formalism results in MHDM charged Yukawa matrices of the form
\begin{equation}
\label{eq:YukAlign}
Y^{d}_{k} = \zeta^{d}_{k} \,Y^{d}_1, \,\,\,\,\, Y^{u}_{k} = \zeta^{u}_{k} \,Y^{u}_1, \,\,\,\,\, Y^{e}_{k} = \zeta^{e}_{k} \, Y^{e}_1,
\end{equation}
where the $\zeta^{A}$ are complex $3 \times 3$ diagonal matrices:
\begin{equation}
\label{eq:zeta}
\zeta^{A}_{k} = \text{diag} \left(\zeta_{k}^{A_1},\zeta_{k}^{A_2},\zeta_{k}^{A_3}\right)\,,\,\,\,\,\,\, \zeta^{A}_{1} \equiv \mathbb{1} \,,
\end{equation}
and the indices $A_i$ correspond to different fermion generations (e.g. $e$, $\mu$, $\tau$).  \eqref{eq:YukAlign}-\eqref{eq:zeta} are equivalent to eqs.(20-21) of \cite{Penuelas:2017ikk}, which was recognized as a generalised form of the Yukawa Alignment suggested in \cite{Pich:2009sp}, where the matrices $\zeta^{A}_{k}$ were instead limited to be complex numbers (already in the weak-eigenstate basis).  We see that this powerful, tree-level FCNC-free ansatz is therefore naturally realised from RFS dynamics.

We note also that the alignments we obtain in this way are a restricted version of the Cheng-Sher ans\"atze \cite{Cheng:1987rs}, in that we obtain the same textures for all $Y_k$, with the caveat that $Y_1$ is the same texture with additional zeroes, given that it is diagonal in the fermion mass basis. This applies even if we are not in the Higgs basis.

\item {\bf{Renormalisation Group Stability:}}
Since $\mathcal{G}_A$ are not symmetries of the full MHDM Lagrangian one should not see them as protecting against quantum corrections in the same way that $\mathbb{Z}_n$ symmetries do in models with natural flavour conservation or special Yukawa textures \cite{Glashow:1976nt,Paschos:1976ay,Weinberg:1976hu,Branco:2011iw}.  Indeed, $\mathcal{G}_A$ appear only after the full symmetry $\mathcal{G_{F,Q,L}}$ is \emph{broken}.\footnote{It may be possible to append protecting symmetries to $\mathcal{G_{F,Q,L}}$ which are left unbroken after flavons develop VEV, but this is beyond our discussion.} One must therefore consider the renormalisation group evolution (RGE) of MHDM Yukawa couplings, which are known at one-loop order \cite{Ferreira:2010xe}.

As FCNC are already permitted at tree level in the weak RFS limit we will not discuss further radiative effects there.  In the strict RFS limit and for the special case of `normal' alignment \cite{Pich:2009sp}, where $\zeta^A$ are complex numbers (not matrices), \cite{Ferreira:2010xe} found that the shape is not stable under the one-loop RGE unless certain relationships between $\zeta^A$ hold, which project onto the naturally flavour conserving variants of MHDM, e.g. the type-I, -II, -X, and -Y 2HDM.  \cite{Botella:2015yfa} subsequently showed that when the $\zeta^A$ parameters are also allowed to evolve under the RGE (and not just the Yukawas, in the weak-eigenstate basis), then special textures like the `democratic' matrix of \cite{Branco:1990fj,Fritzsch:1994yj} are also aligned across all scales. As mentioned, all of these special cases can be enforced, and therefore the alignment can be RGE stable, by imposing simple Abelian discrete symmetries.

However, some misalignment is phenomenologically acceptable and even interesting.  The authors of \cite{Penuelas:2017ikk} have shown that the leading-order operator $\mathcal{L}_{FCNC}$ induced at one-loop in the generalised alignment scenario \eqref{eq:YukAlign} is proportional to (in the Higgs and neutral scalar mass-eigenstate basis)
\begin{align}
\nonumber
\tilde{\Theta}^d_k = &- V^{\dagger}\,\sum_{l=1}^{N}\, \zeta^{u,\dagger}_l \,m_u m^{\dagger}_u \,\zeta^u_k \, V \, \zeta^d_l \\
\nonumber
&+  \zeta^{d}_k \, V^{\dagger}\,\sum_{l=1}^{N}\, \zeta^{u,\dagger}_l \,m_u m^{\dagger}_u \, V \, \zeta^d_l \\
\label{eq:FCNC1loop}
&+ \frac{1}{4}\left[ V^{\dagger} \left(\sum_{l=1}^{N} \zeta^{u,\dagger}_l m_u m_u^\dagger \zeta^u_l \right)V, \,\,\,\zeta^d_k  \right] \,,
\end{align}
where the last term is a commutator. A similar expression holds for the up quarks, while the RGE for charged leptons does not misalign. The term in \eqref{eq:FCNC1loop} couples to neutral Higgs fields, left- and right-chiral down quarks, and an additional power of $m_d$.  That is, the leading FCNC contributions are dimension seven and are suppressed by two CKM factors, three alignment factors $\zeta^A$, and three mass insertions.  This structure can be understood in terms of the low-energy re-phasing freedoms/symmetries of the SM fermions and Yukawa, mass, and mixing matrices.  

It is found in \cite{Penuelas:2017ikk} that significant constraints on the $\zeta^A$ parameters can be obtained from $\bar{B}^0_s \rightarrow \mu^+\mu^-$ and $\bar{B}^0_s- B^0_s$ mixing observables in the 2HDM and normal alignment cases, and also that Yukawa Alignment at very high scales satisfies all experimental bounds (which is consistent with the idea of an RFS appearing at the flavour-breaking scale).
\end{enumerate}

\section{Sketching a Realistic Toy Model}
\label{sec:MODEL}

In this section we consider an explicit model, similar to the toy models of \cite{Varzielas:2011jr} (in that a specific flavon connects to a given type of fermion), that realises the RFS mechanism in an MHDM. We take as the example a simple but viable $A_4$ family symmetry model addressing just the lepton sector, which represents a variation of the Altarelli-Feruglio model discussed in Section \ref{sec:REC} that is both UV complete \cite{Varzielas:2010mp} and addresses non-zero $\theta^{l}_{13}$ \cite{Varzielas:2012ai}. Then we simply add multiple Higgs which transform trivially under $A_4$, such that the Yukawa couplings to any of the Higgs are all controlled by the same family symmetry breaking VEV. 

The analysis for the charged leptons in \cite{Varzielas:2012ai} proceeds analogously to Altarelli-Feruglio, in that only the $A_4$ triplet flavon $\phi_l$ couples to create the Yukawa couplings, and again leaves invariant the $\mathbb{Z}_3$ subgroup of $A_4$ after obtaining its VEV given in \eqref{eq:flavonvev}.\footnote{In this toy model we do not explicitly assume supersymmetry, but instead take the VEV as given.  We also do not discuss further shaping or Froggatt-Nielsen symmetries irrelevant to our discussion.} Conveniently, in the symmetry basis we are working in, this VEV direction and the representations of the RH charged leptons from Table \ref{tab:AF} again lead to diagonal charged lepton Yukawa couplings.

The key point here is that this still applies when adding multiple Higgs. We re-label $H^{\prime}$ as $H^{\prime}_1$ and explicitly add $H^{\prime}_2$ (it will be clear that adding further $H_{k}^{\prime}$ does not affect the mechanism). The symmetry invariants for the charged lepton terms can be written as:
\begin{align}
\nonumber
\frac{H^{\prime}_1}{\Lambda} & \left( y_e^1 [ \bar{L}_L \phi_l ] e_R + y_\mu^1 [ \bar{L}_L \phi_l ]' \mu_R + y_\tau^1 [ \bar{L}_L \phi_l ]'' \tau_R \right) + \\
\frac{H^{\prime}_2}{\Lambda} & \left( y_e^2 [ \bar{L}_L \phi_l ] e_R + y_\mu^2 [ \bar{L}_L \phi_l ]' \mu_R + y_\tau^2 [ \bar{L}_L \phi_l ]'' \tau_R \right)\,.
\end{align}
The symmetry doesn't control the dimensionless couplings $y$, which can easily originate from renormalisable terms in the UV completion \cite{Varzielas:2010mp, Varzielas:2012ai} that could also relate the scale $\Lambda$ with the mass of some heavy fermion messengers.

This is an explicit multi-Higgs model realising the RFS picture described in Section \ref{sec:REC}, as the Yukawa matrices obtained from the terms above have their structure entirely determined by the $A_4$ symmetry and the direction of $\langle \phi_l \rangle$, and they are diagonal in the same basis (which conveniently is the symmetry basis we are working in):
\begin{align}
\label{eq:sampleYk}
Y_1^{e,\prime} = \frac{v_l}{\Lambda}
\begin{pmatrix}
y_e^1 & 0 & 0 \\
0 & y_\mu^1 & 0 \\
0 & 0 & y_\tau^1
\end{pmatrix}, \,\,
Y_2^{e,\prime} = \frac{v_l}{\Lambda}
\begin{pmatrix}
y_e^2 & 0 & 0 \\
0 & y_\mu^2 & 0 \\
0 & 0 & y_\tau^2
\end{pmatrix}.
\end{align}
One can go to the Higgs basis without loss of generality by performing a unitary transformation in $H^{\prime}_1$ and $H^{\prime}_2$ such that all the VEV resides in $H_1$ --- this merely redefines the relation between the $y$ couplings and the charged lepton masses. We see then that \eqref{eq:sampleYk} corresponds to the Generalised Yukawa Alignment of {\bf{(A6)}}.

The Dirac neutrino masses in the MHDM framework are analogous to those in \eqref{eq:AFSP},
\begin{align}
\label{eq:diracmass}
H^{\prime}_1\, y_{\nu}^{1} \left[\bar{L}_L\nu_R \right] + H^{\prime}_2 \, y_{\nu}^{2} \left[\bar{L}_L \nu_R \right]\,,
\end{align} 
and the Majorana masses given by
\begin{align}
\label{eq:majoranamass}
\left(x_A \xi + x_A^{\prime} \xi^{\prime} \right) \left[\nu_R\nu_R \right] + x_B \left[\phi_\nu\nu_R\nu_R\right]
\end{align}
similarly mimic \eqref{eq:AFSP}, although here the flavon $\xi'$ is a \emph{non}-trivial ($1^{\prime}$) $A_4$ singlet, and is ultimately responsible for generating the non-zero $\theta_{13}^{l}$ \cite{Varzielas:2012ai}.  Upon application of the Type-I seesaw mechanism the additional MHDM mass term from \eqref{eq:diracmass} only changes the overall normalization of the low-energy Majorana mass matrix from \cite{Varzielas:2012ai}, and so the RFS analysis presented there still holds --- the $\xi^{\prime}$ breaks the original Altarelli-Feruglio $\mathbb{Z}_2 \times \mathbb{Z}^{\prime}_2$ RFS preserved by $\langle \phi_\nu \rangle$, with $\mathbb{Z}^{\prime}_2$ an accidental $\mu-\tau$ operator enforcing $\theta^{l}_{13} = 0$, down to a single $\mathbb{Z}_2$ generated by
\begin{equation}
\label{eq:TnuAF2}
T_{\nu U} = U_{TM} \cdot \,\underbrace{\text{diag}(-1,1,-1)}_{T_{\nu}} \cdot \,U_{TM}^{\dagger} \,,
\end{equation}
in the flavour basis.  Unlike the $\mu-\tau$ operator generating $\mathbb{Z}^{\prime}_2$, this remaining $\mathbb{Z}_2$ is an $A_4$ subgroup, and therefore a proper RFS.  In \eqref{eq:TnuAF2}, $U_{TM} \equiv U_{TBM} \cdot R^{13}(\delta_{13}=0)$ and, as per our discussion below \eqref{eq:mixdegen}, this model allows a free mixing parameter $\theta$ thanks to the lone invariance under \eqref{eq:TnuAF2}.  The parameter $\theta$ can be fit from the model's prediction for the physical leptonic reactor angle $\theta_{13}^{l}$,
\begin{equation}
| \sin \theta_{13}^{l} | = \frac{2}{\sqrt{6}} | \sin \theta | \,,
\end{equation}
and then a phenomenologically viable PMNS matrix is obtained.  

As a final point we observe that we avoid the `no-go' theorem of \cite{Felipe:2014zka}, since we take the Higgs fields to be $A_4$ singlets (not triplets), and further include triplet flavons $\phi_{l,\nu}$, both of which break $A_4$.  This differs from the setups of \cite{Felipe:2014zka,Felipe:2013ie,Felipe:2013vwa,Pramanick:2017wry}, for example.
\section{Conclusion}
\label{sec:CONCLUDE}

We have imposed RFS on the Yukawa sector of MHDM and shown how, in the limit where the RFS controls all parameters of the three-generation CKM and PMNS matrices, the Yukawa couplings for each $H_k$ are strictly diagonal, and therefore free of tree-level FCNC.  This limit corresponds to a generalised form of Yukawa Alignment \cite{Pich:2009sp,Penuelas:2017ikk}.  In weaker limits when only portions of the CKM/PMNS are controlled by the RFS, the tree-level FCNC are limited to flavour-specific patterns that can still avoid experimental bounds.  Our results are model-independent, so long as one assumes family symmetry breaking patterns similar to those in \eqref{eq:GF}.  In addition to discussing various subtleties of our approach, including the implications of RGE between disparate scales, we have also sketched an explicit Altarelli-Feruglio inspired $A_4$ model that exhibits the RFS, predicts a viable PMNS mixing matrix, and yields the Yukawa Alignment, in order to show how our results can be achieved in realistic BSM setups.  Our approach is therefore model-inspired, but does not rely on the dynamics of any particular UV Lagrangian. 

\section*{Acknowledgements}

We thank Lu\'is Lavoura for helpful conversations and Jo\~ao P. Silva for a review of the manuscript.
IdMV acknowledges
funding from Funda\c{c}\~{a}o para a Ci\^{e}ncia e a Tecnologia (FCT) through the
contract IF/00816/2015 and was supported in part by the National Science Center, Poland, through the HARMONIA project under contract UMO-2015/18/M/ST2/00518 (2016-2019), and by Funda\c{c}\~ao para a Ci\^encia e a Tecnologia (FCT,
Portugal) through the project CFTP-FCT Unit 777 (UID/FIS/00777/2013) which is partially funded through POCTI (FEDER), COMPETE, QREN and EU.  J.T. acknowledges funding from DESY and thanks the CFTP in Lisbon for its hospitality and support while portions of this project were completed.

\end{document}